\documentclass[10pt,twocolumn,english,conference]{IEEEtran}
\usepackage{multicol}
\usepackage{amsmath}
\usepackage{subeqnarray}
\usepackage{cases}
\usepackage{framed}
\usepackage{amsfonts}
\usepackage{amssymb}
\usepackage{bm}
\usepackage{graphicx}
\usepackage{subfigure}
\usepackage{babel}
\usepackage{cite}
\usepackage{color}%
\usepackage{algorithm}

\providecommand{\U}[1]{\protect\rule{.1in}{.1in}}

\newcommand{\tr}{\mathsf{Tr}}
\newcommand{\rank}{\mathsf{Rank}}

\begin{document}

\title{Using Wireless Network Coding to \\
Replace a Wired with Wireless Backhaul}
\author{%

\begin{tabular}
[c]{c}%
Henning Thomsen, Elisabeth de Carvalho, Petar Popovski\\
Department of Electronic Systems, Aalborg University, Denmark\\
Email:  $\left\{ {{\text{ht,edc,petarp}}} \right\}$@es.aau.dk\\
\end{tabular}
}
\maketitle

\begin{abstract}
Cellular networks are evolving towards dense deployment of small cells. This in turn demands flexible and efficient backhauling solutions. A viable solution that reuses the same spectrum is wireless backhaul where the Small Base Station (SBS) acts as a relay. In this paper we consider a reference system that uses wired backhaul and each Mobile Station (MS) in the small cell has its uplink and downlink rates defined. The central question is: if we remove the wired backhaul, how much extra power should the wireless backhaul use in order to support the same uplink/downlink rates? We introduce the idea of wireless-emulated wire (WEW), based on two-way relaying and network coding. Furthermore, in a scenario where two SBSs are served simultaneously, WEW gives rise to new communication strategies, partially inspired by the private/public messages from the Han-Kobayashi scheme for interference channel. We formulate and solve the associated optimization problems. The proposed approach provides a convincing argument that two-way communication is the proper context to design and optimize wireless backhauling solutions. 

\end{abstract}

\section{Introduction}\label{sec:Introduction}
The next leap increasing the wireless data rates for multiple users is bringing the access point closer to the users and enable spatial reuse over smaller distances. This is often referred to as \emph{network densification}~\cite{Qualcomm5G} and is seen as a key feature of the upcoming 5G wireless systems. The trend of densification has already started with the deployment of femtocells (small cells) and creation of heterogeneous networks~\cite{andrews2012femtocells}, and small cells are seen as the key architectural feature in the very dense networks that will be operational within the next decade. 

The proliferation of small cells is conditioned on the existence of a flexible, cost-effective backhaul solution~\cite{andrews2013seven} between a Base Station (BS) and a Small-cell Base Station (SBS), which can permit rapid deployment of SBSs to meet the demand of the increasing number of data-hungry users. From the viewpoint of wireless performance and spectrum usage, wired backhaul with copper or fiber-optical links is certainly a desirable solution, as the backhauled data does not create additional interference or use extra spectrum resources. Recently, methods based on millimeter wave have also been investigated as enablers of wireless backhaul~\cite{6600706}, motivated by the ample spectrum available in the 60 GHz region. However, this would represent a wireless backhaul solution with two coexisting radio interfaces at the SBS. 

Here we use the term wireless backhaul for a system in which the link BS-SBS uses the same spectrum as the link Mobile Station (MS)-SBS, such that SBS has effectively the role of an in-band relay. During the last decade, a relay in the context of wireless cellular networks has been seen as an enabler of improved coverage~\cite{DohlerCommMag}. However, the critics have repeatedly pointed out its inherent loss of spectral efficiency and, until now, it has still not become a part of the architectural mainstream in the area of wireless cellular networks. Wireless Network Coding (WNC) has introduced a fresh potential in the area of relaying, by exhibiting significant gains in spectral efficiency for scenarios with two-way relaying~\cite{Petar06,Petar07,PLNC}. Perhaps the most important message from WNC is that if wireless communication problems are defined by considering two-way traffic from the terminals, rather than the traditional decoupling of the uplink and downlink, then the space of possible communication strategies and potential gains is largely expanded. This has been demonstrated, for example, in~\cite{ChanTCOM}, \cite{HuapingWCL} and \cite{liu2013mimo}, where the definition of a two-way wireless problem lead to schemes such as Coordinated Direct and Relay (CDR) in~\cite{ChanTCOM} and four-way relaying for single-antenna nodes in~\cite{HuapingWCL}, and multiple-antenna nodes in~\cite{liu2013mimo}. 

%
%

\begin{figure}[t!]
	\centering
		\includegraphics[width=7cm]{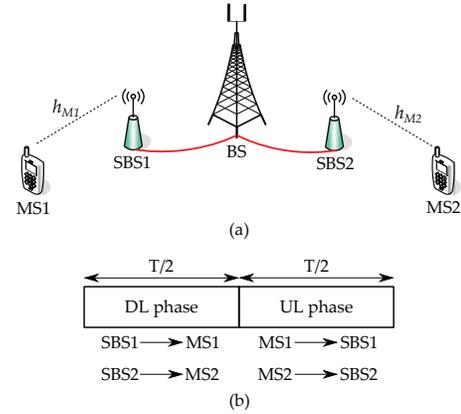}
	\caption{(a) A reference system with two small cells. (b) The transmissions in the uplink and downlink; only the wireless transmissions are depicted, the transmissions over the wired backhaul are taking place in parallel.}
	\label{fig:SystemModelWired}
\end{figure}
\begin{figure}
	\centering
		\includegraphics[width=7cm]{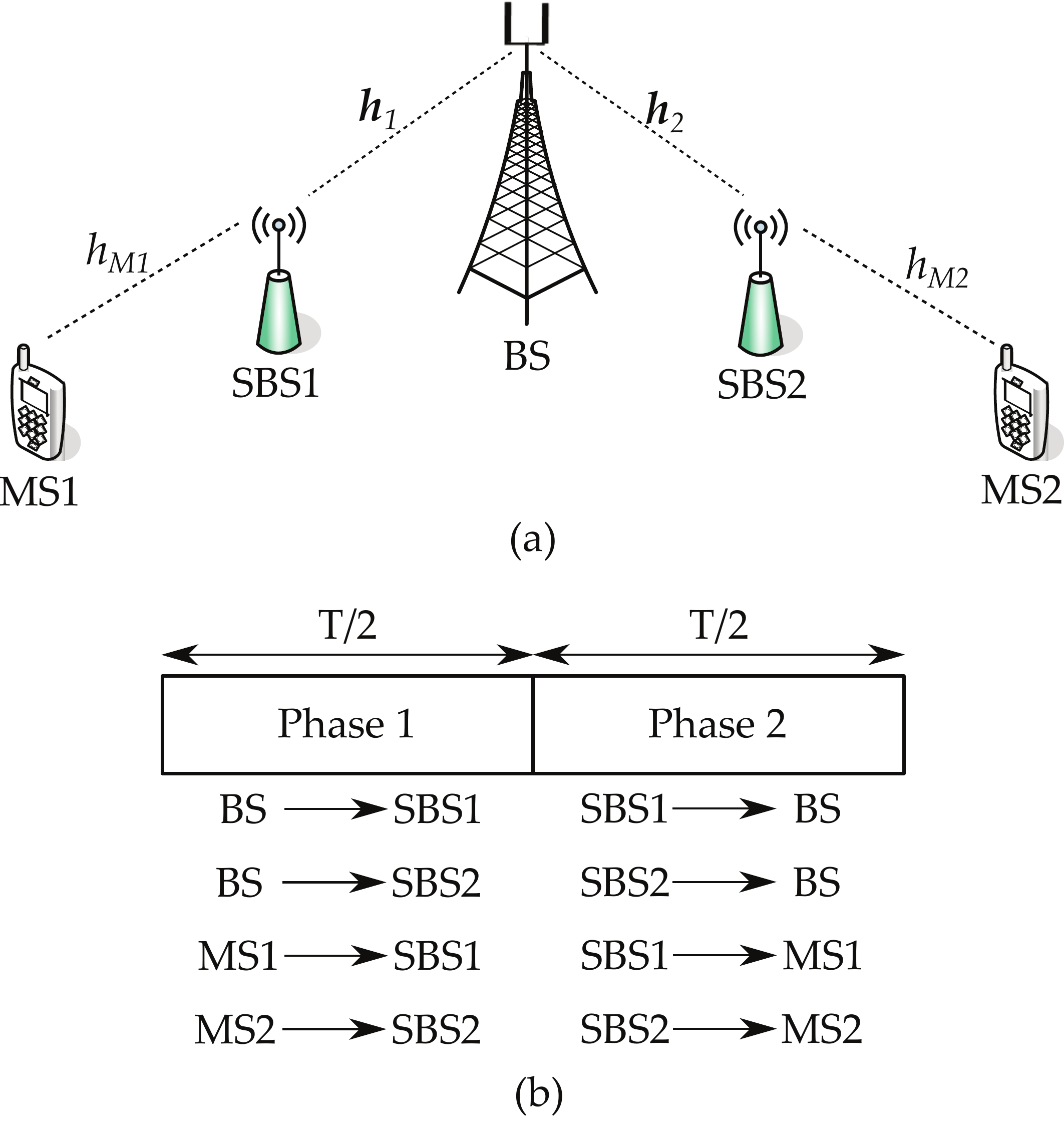}
	\caption{System Model used in this paper, where the two SBSs are antipodal, and each of them serve two-way traffic flow of the MS.}
	\label{fig:SystemModel}
\end{figure}
The problem definition in this paper can be explained by using Figs.~\ref{fig:SystemModelWired} and~\ref{fig:SystemModel}. 
Fig.~\ref{fig:SystemModelWired} shows the reference system, with a macro BS and two small cells, but for this initial explanation we ignore SBS2 and MS2 and focus only on SBS1 and MS1. The system is based on Time Division Duplex (TDD), where the elementary time unit has a duration $T$ and, within that time, an equal time of $\frac{T}{2}$ is allocated to the Downlink (DL) and the uplink (UL) transmission, respectively. The DL/UL transmission rates of MS$1$ are $R_{D1}/R_{U1}$, respectively and these rates are perfectly supported by the backhaul. Note that the DL,UL rates when observed over the whole interval $T$ are $\left( \frac{R_{D1}}{2}, \frac{R_{U1}}{2}\right)$. The central question is: \emph{Can we remove the wired backhaul and still support the same rate pair $\left( \frac{R_{D1}}{2}, \frac{R_{U1}}{2}\right)$ within the interval of length $T$, without requiring any changes in the baseband of MS1?}

The answer is affirmative and depicted on Fig.~\ref{fig:SystemModel}. We use the basic idea of two-way relaying with WNC and devise transmission in two phases. In Phase 1, BS and MS1 transmit simultaneously. MS1 transmits with the same power and rate $R_{U1}$, as in Fig.~\ref{fig:SystemModelWired}. BS now need to use power $P$ for wireless transmission, while it did not use any on Fig.~\ref{fig:SystemModelWired}, and it transmits at a rate $R_{D1}$. Assuming that $R_{U1}$ is equal to the capacity of the link MS1-SBS1, it follows that SBS1 must be able to decode the ``clean'' signal of MS1, without any residual interference from the transmission of BS. Therefore, the minimal $P$ selected by the BS should allow SBS1 to decode the signal of rate $R_{D1}$ from the BS by treating the signal from MS1 as a noise, then cancel the signal form the BS and proceed to decode the signal of rate $R_{U1}$ from MS1. In Phase 2, SBS1 uses WNC by XOR-ing the decoded messages and broadcasts them to MS1 and BS, respectively. Thus, at the end of the interval $T$, the performance is equivalent to the one of the wired backhaul and we can say that we have obtained a \emph{wireless-emulated wire (WEW)}.    

In the paper we show how the ideas of WEW can be put to work when the BS needs to serve multiple SBSs, as in the examples on Figs.~\ref{fig:SystemModelWired} and~\ref{fig:SystemModel}. The condition imposed to the SBS$i$ to be able to decode a ``clean'' uplink signal from the respective MS$i$ leads to novel communication strategies in Phase 1, partially inspired by the framework of private and public messages in the Han-Kobayashi schemes for interference channel~\cite{HK}. We introduce these strategies in a setting in which BS has multiple antennas and solve the associated optimization problems, evaluating the minimal required power at the BS and the SBSs to achieve emulation of the wired backhaul.
%

This paper is organized as follows. In the next section, we give the system model and state the assumptions. The proposed scheme is described in detail in Sec.~\ref{subsec:WEW}, and the associated optimization problems  are formulated and solved in Sec.~\ref{sec:OptimizationProblem}. Sec.~\ref{sec:NumericalResults} gives numerical examples and comparisons of our scheme, and the paper is concluded in Sec.~\ref{sec:Conclusion}.

\section{System Model}\label{sec:SystemModel}
%
%
\begin{figure*}
	\centering
		\includegraphics[width=\textwidth]{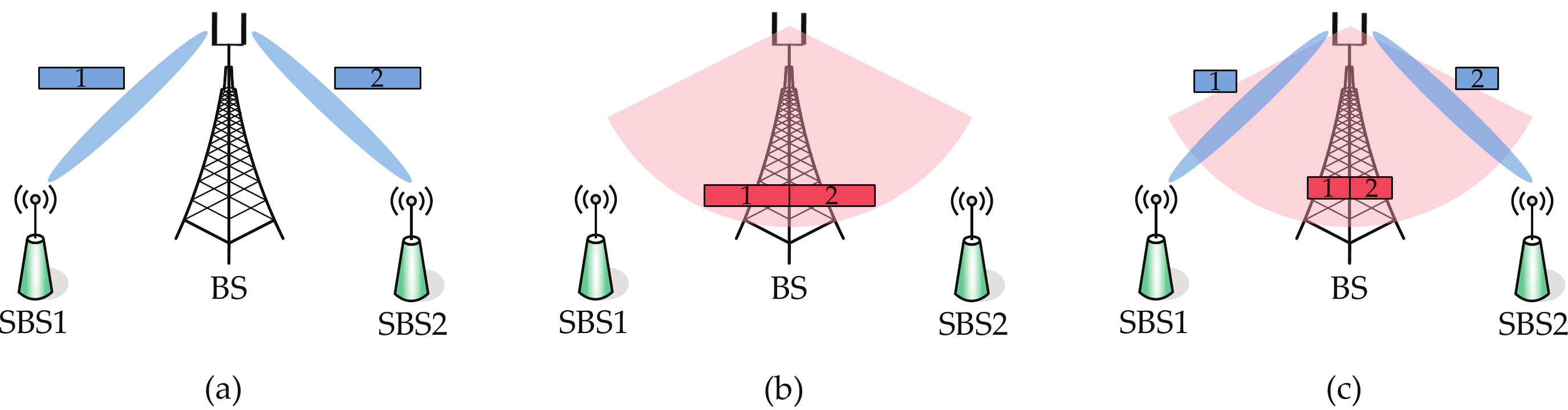}
	\caption{The three transmission methods considered in this paper. In (a), the BS transmits to the SBSs using ZF. In (b), the BS broadcasts the concatenated data to both SBSs. The WEW is shown in (c), wherein the BS transmits data using both ZF, and the common beamforming.}
	\label{fig:TransmissionAllThreeMethods}
\end{figure*}
\subsection{Reference System: Wired Backhaul}

The reference system model considered is a setup with a \emph{wired} backhaul, as shown in Fig.~\ref{fig:SystemModelWired}(a), with one macro BS and 2 SBS. One MS is attached to each SBS. 
Each MS$i$ has both uplink and downlink data to and from the BS, which transit through the corresponding SBS. 
Both SBSs and MSs are equipped with a single antenna. 
Each MS$i$ has a wireless connection to its SBS$i$ through the channel $h_{Mi}$. These channels are assumed to be reciprocal and remain constant over the duration of  two transmission phases. 
The SBS$i$ and BS are connected by wire, as shown by the red link in Fig.~\ref{fig:SystemModelWired}(a). The wired connection is assumed to have a sufficient capacity to support the uplink and downlink rate requirements of the MSs.
The SBS$i$ transmits at power $P_{Si}$, and the MS$i$ at power $P_{Mi}$, and it is assumed that $P_{Mi} \le P_{Si}$. In this paper, we do not consider the case when outage occurs. Thus we assume that both the uplink rate $R_{Ui}$ and downlink rate $R_{Di}$ of each MS are selected to be the highest possible:
\begin{align}
	R_{Ui} &= \log_2 \left( 1 + \frac{P_{Mi} \vert h_{Mi} \vert^2}{\sigma^2} \right) \; \text{ and } \label{eqn:MSuplinkRateWired} \\
	R_{Di} &= \log_2 \left( 1 + \frac{P_{Si} \vert h_{Mi} \vert^2}{\sigma^2} \right) \label{eqn:MSdownlinkRateWired}
\end{align}hold, where $\sigma^2$ is the power of the Additive White Gaussian Noise (AWGN). Note that as a consequence of the assumption $P_{Mi} \le P_{Si}$, and of channel reciprocity, we have $R_{Ui} \le R_{Di}$. 
\subsection{Wireless Backhaul}
In the system under study,  a wireless backhaul replaces the wired backhaul, as shown in Fig.~\ref{fig:SystemModel}(a). The assumptions considered  are the same as the ones described for the wired backhaul, except for the assumptions involving the wireless link between BS and the SBSs. Transmission of the uplink and downlink traffic is still performed in two phases as explained in section~\ref{subsec:WEW}. 
All nodes are half-duplex, so they cannot transmit and receive at the same time. 
The SBSs and BS use the same spectrum, such that SBS acts as a relay. The SBSs operate in Decode-and-Forward (DF) mode, and full Channel State Information (CSI) is assumed available at all nodes.

The BS has $2M$ antennas and the wireless backhaul link between SBS$i$ and BS is modeled as the channel vector $\bm{h}_i \in \mathbb{C}^{\left[2M \times 1\right]}$, where $\mathbb{C}$ is the complex numbers. These channels are assumed reciprocal. 
Furthermore, we assume the small cells are assumed to be sufficiently separated, such that there is no interference between them. 

The transmission power of MS$i$  is identical to the wired backhaul case. 
However, due to the use of the channel $\bm{h}_i$, not used in the wired backhaul case, the transmission power $P_{Si}$ used by SBS$i$ in the wired case may not be sufficient to ensure an outage-free transmission to the BS, such that it may need to be increased, adding to the total power budget required for wireless backhauling. 
\section{Wireless-Emulated Wire}\label{subsec:WEW}
%
The two key ingredients of the Wireless-Emulated Wire (WEW) are the use of WNC at each SBS and message splitting at the BS. 
\subsection{Transmission based on WNC}
Transmission of the uplink and downlink traffic is performed in two equal-duration phases, as for the wired counterpart, since the goal is to exactly emulate the wired backhaul system. 
The two phases, phase 1 and phase 2, are shown in Fig.~\ref{fig:SystemModel}(b).Transmission occurs as follows:
\begin{itemize}
	\item In phase 1, each MS$i$ transmits its uplink message at rate $R_{Ui}$ to SBS$i$. Simultaneously, BS transmits the downlink messages to both SBSs at rates $R_{D1}$ and $R_{D2}$.
	\item In phase 2, each SBS$i$ decodes all received messages, and broadcasts the XOR of them to both SBS$i$ and BS. The BS/each SBS$i$ decode the received message and XOR it with the message transmitted previously, thus obtaining the desired message from each SBS$i$/BS, respectively. 
\end{itemize}
The details of phase 1 are given in Section~\ref{subsec:MAphase}, and of phase 2 in Section.~\ref{subsec:BCphase}.

Note that even though the first phase resembles the Multiple Access phase of traditional two-way relaying, the proper way to look at this is as a constrained broadcast phase, as seen from the BS point of view. The constraint is that, at each SBS, the signal from the MS$i$ be decoded interference-free, meaning that downlink signals from the BS should be decoded first and their contribution removed from the received signal before proceeding to the decoding of the signal from the MS. 
An important aspect of the transmission optimization is that the transmit power from the BS should be adjusted such that the downlink signals be decoded at the SBS considering the uplink signal as interference. This decoding procedure guarantees the MS can send at rate $R_U$, equal to the single-user capacity, while maintaining the same power for the MS. 
%

\subsection{Private and Common Messages}

Our scheme is designed to benefit from two transmission options at the BS. 
A natural choice is to use Zero Forcing (ZF) beamforming at the BS and send data to the two SBSs through two orthogonal spatial channels. Each SBS$i$ receives only  its intended message from the BS. This is shown in Fig.~\ref{fig:TransmissionAllThreeMethods}(a), where message 1 is sent to SBS1, and message 2 to SBS2. Because the message sent by ZF is received only by the intended SBS, it is referred to as a \emph{private} message. 

The use of ZF beamforming is detrimental when the channel is ill-conditioned, resulting in noise enhancement at low SNR. This situation will occur when uplink rate is equal or larger than the downlink rate. 
MMSE  beamforming is usually seen as an viable alternative. However, MMSE beamforming cannot be used as it leaves a residual interference at the SBS that cannot be decoded, hence violating our condition that MS should be able to send to SBS over a channel identical as with the wired backhaul. 
In the extreme case where the channels  $\bm{h}_1$ and  $\bm{h}_2$ are collinear, a better solution is to send a  \emph{common} message that is broadcasted to both receivers, as shown in Fig.~\ref{fig:TransmissionAllThreeMethods}(b). 
Here, the two messages (data bits) are concatenated at the BS into a \emph{common message}, which is then encoded/transmitted. Both SBSs must decode this common message it in its entirety, and therefore remove its contribution before decoding the uplink signal. 

In WEW, transmission of the BS mixes private and common message, as shown in Fig.~\ref{fig:TransmissionAllThreeMethods}(c). The BS splits the message for each MS into two parts, a private part and a common part. 
The private part is sent using ZF. The common parts of the messages to both MSs are concatenated, and transmitted using a common beamformer. SBS1 receives the private message 1 (coloured blue), and the common message (coloured red). It must then decode both the private message and the common message. Similarly, SBS2 decodes its private message and the common message. After decoding, SBS$i$, $i=1,2$ extracts from the common message the data bits intended for MS$i$ and recreates the original message intended from BS to MS$i$, before XOR-ing it with the message received from MS$i$ and broadcasting the XOR-ed message to BS and MS$i$. 

Note that in the figure, the size of the private and common message for each MS are equal. 
In practice, the ratio between private and common bit length is determined in the proposed optimization procedure. 
The cases in Fig.~\ref{fig:TransmissionAllThreeMethods}(a)-(b) are special cases of Fig.~\ref{fig:TransmissionAllThreeMethods}(c).

\subsection{Two-phase Transmission}

\subsubsection{Phase 1}\label{subsec:MAphase}
The BS transmits the downlink message, intended for MS$i$, to SBS$i$ at rate $R_{Di}$. The message is split into two parts, the private and common part, which are sent at rates
\begin{equation}\label{eqn:ratealpha}
	R_{Pi} = \alpha_i R_{Di}, \hspace{0.5cm} R_{Ci} = \left( 1 - \alpha_i \right) R_{Di}
\end{equation}
respectively. Here $0 \le \alpha_i \le 1$, $i=1,2$. So $\alpha_i$ is the fraction of the message to be transmitted privately, and $1 - \alpha_i$ the fraction to be sent publicly. Thus $R_{Di} = R_{Pi} + R_{Ci}$.

Physically, BS transmits the signal
\begin{equation}\label{eqn:transmittedSignal}
	 \sqrt{P_1}\bm{w}_1 x_1 + \sqrt{P_2}\bm{w}_2 x_2 + \bm{w}_C x_C,
\end{equation}
where $x_1$, $x_2$ is the data sent privately, for user 1 and 2, respectively, and $x_C$ is the common data. The private data is sent using transmission power $P_1$ and $P_2$ respectively. The beamformers $\bm{w}_1, \bm{w}_2 \in \mathbb{C}^{\left[ 2M \times 1\right]}$ are defined using the Zero Forcing (ZF) condition, i.e. they must satisfy
\begin{equation}\label{eqn:ZFcondition}
	\bm{h}_1^H \bm{w}_2 = 0 \; \text{ and } \; \bm{h}_2^H \bm{w}_1 = 0.
\end{equation}
For the 2 stream case, these two beamformers are given by~\cite{brown2012practical}
\begin{equation}\label{eqn:ZFbeamformers}
	\bm{w}_i = \left( I_{2M} - \bm{h}_j \left( \bm{h}_j^H\bm{h}_j\right)^{-1} \bm{h}_j^H\right)\bm{h}_i,
\end{equation}
for $i,j=1,2$, $i \neq j$, $I_{2M}$ is the $2M \times 2M$ identity matrix and $(\cdot)^H$ is Hermitian transpose. For analytical convenience, the ZF beamforming vectors are normalised, so they have unit norm. They are then scaled with the transmission powers $\sqrt{P_1}$ and $\sqrt{P_2}$. The beamformer $\bm{w}_C \in \mathbb{C}^{\left[ 2M \times 1\right]}$ is used for transmission of the common message, sent at power $P_C = \Vert \bm{w}_C \Vert^2$, and is determined based on the optimization procedure described in Sec.~\ref{sec:OptimizationProblem}. Here $\Vert \bm{w}_C \Vert$ is the euclidean norm of $\bm{w}_C$.

Simultaneously, MS$i$ transmits $x_{Mi}$ at rate $R_{Ui}$ through the channel $h_{Mi}$ to SBS$i$. Using~\eqref{eqn:MSdownlinkRateWired}, and defining $\gamma_{Mi} = \frac{P_{Mi}\vert h_{Mi} \vert^2}{\sigma^2}$, this rate equals
\begin{equation}\label{eqn:userUplinkCapacity}
	R_{Ui} = \log_2 \left( 1 + \gamma_{Mi} \right).
\end{equation}


The signal received at SBS$i$ is
\begin{align}\label{eqn:signalSBS$i$}
	y_{Si} =& \bm{h}_i^H \left( \sqrt{P}_1\bm{w}_1 x_1 + \sqrt{P}_2\bm{w}_2 x_2 + \bm{w}_C x_C \right) \\ 
	&+ h_{Mi}x_{Mi} + z_{Si} \nonumber \\
	=& \bm{h}_i^H \sqrt{P}_i\bm{w}_i x_i + \bm{h}_i^H \bm{w}_C x_C + h_{Mi}x_{Mi} + z_{Si},
\end{align}
where we have used the ZF condition in~\eqref{eqn:ZFcondition}, and where $z_{Si}$ is the AWGN at SBS$i$.
Each SBS is required to decode all received signals with the rate requirements previously mentioned. 
At each SBS$i$, we have a Multiple Access Channel (MAC) rate region, dependent on the transmission powers $P_i$ and $P_C$.

\subsubsection{Phase 2}\label{subsec:BCphase}
After each SBS$i$ has decoded the private and the common message, it first recreates the original message intended for MS$i$, computes the XOR of that message and the message decoded from the MS$i$ and broadcasts. By assumption, the SBS-MS link can support the uplink and downlink rates, since we assume this channel to be identical to the wired backhaul case. Then~\eqref{eqn:MSuplinkRateWired} and~\eqref{eqn:MSdownlinkRateWired} also hold in the wireless backhaul case.
As mentioned earlier in this section, the SBS$i$-BS link may not be able to support the rates, so the transmission power at each SBS$i$ may need to be increased. We deal with this problem in Subsec.~\ref{sec:transPowerSBS}.

At the end of phase 2, BS, MS1 and MS2 decode the signal sent by their respective SBS$i$.As in standard in network coding, each end node uses XOR to recover the desired message from the network-coded message. 

\section{Optimization Problems}\label{sec:OptimizationProblem}

In this section, we define the two optimization problems involved in WEW. The first problem is finding the minimal transmission power at the BS, subject to the given rate constraints, along with the split factors between common and private messages at both SBSs. The second one is finding the additional power needed so that the signal broadcast by SBS is reliably received by BS and the MS.

\subsection{Optimizing the Transmission Power and Private/Common Message a the BS }\label{subsec:transPowerBS}

Here we minimize the transmit power at the BS, subject to the given uplink and downlink rate constraints. The optimization variables are the transmission powers $P_1$ and $P_2$ for the ZF beamformers, the common beamformer $\bm{w}_C$ (including the associated  transmit power), as well as the splitting factors $\alpha_1$ and $\alpha_2$ between private and common messages. 
Define the SNRs
\begin{equation}\label{eqn:SNRsdef}
	\gamma_{Pi} = \frac{P_i \vert \bm{h}_i^H \bm{w}_i \vert^2}{\sigma^2}, \hspace{0.5cm} \gamma_{Ci} = \frac{\vert \bm{h}_i^H \bm{w}_C \vert^2}{\sigma^2}.
\end{equation}
for $i=1,2$. Also, using~\eqref{eqn:userUplinkCapacity}, the SNR of the MS$i$-SBS$i$ link is $\gamma_{Mi} = 2^{R_{Ui}}-1$. Using this, the optimization problem is stated below.

\begin{equation*}
\begin{aligned}
& \underset{P_i, \bm{w}_C, \alpha_i}{\text{minimize}}
& & P_1 + P_2 + \Vert \bm{w}_C \Vert^2\\
& \text{subject to}
& & R_{Pi} \le \log_2\left( 1 + \frac{\gamma_{Pi}}{1+\gamma_{Mi}} \right) \\
& & & R_{Ci} + R_{Cj} \le \log_2\left( 1 + \frac{\gamma_{Ci}}{1+\gamma_{Mi}} \right) \\
& & & R_{Pi} + R_{Ci} + R_{Cj} \le \log_2\left( 1 + \frac{\gamma_{Pi} + \gamma_{Ci}}{1+\gamma_{Mi}} \right) \\
& & & P_i \ge 0, \;\;0 \le \alpha_i \le 1, i,j=1,2, i \neq j
\end{aligned}
\end{equation*}
The objective function is the sum of the transmission powers $P_1$ and $P_2$ for the two ZF beamformers, and the power $\Vert \bm{w}_C \Vert^2$ in the common beamformer.

As explained in the previous section, we have a MAC rate region at each SBS$i$. The first constraint requires SBS$i$ to decode the private message $x_i$, at rate $R_{Pi}$. The second constraint means that SBS$i$ must decode the common message. The third constraint is the sum-rate constraint at the MAC. Note that in all constraints, we divide by $(1 + \gamma_{Mi})$ in the logarithms. This term corresponds to the decrease in SNR when 
the uplink signal is considered as noise in the decoding of the downlink signals. Let
\begin{align*}
	\beta_{1i} &= \sigma^2\left( 2^{R_{Pi}} - 1 \right)\left( 1 + \gamma_{Mi} \right), \\
	\beta_{2i} &= \sigma^2\left( 2^{R_{Ci} + R_{Cj}} - 1 \right)\left( 1 + \gamma_{Mi} \right), \\
	\beta_{3i} &= \sigma^2\left( 2^{R_{Pi} + R_{Ci} + R_{Cj}} - 1 \right)\left( 1 + \gamma_{Mi} \right).
\end{align*}
Then the problem can be rewritten into the following form
\begin{equation*}
\begin{aligned}
& \underset{P_i, \bm{w}_C, \alpha_i}{\text{minimize}}
& & P_1 + P_2 + \Vert \bm{w}_C \Vert^2\\
& \text{subject to}
& & \beta_{1i} \le P_{i} \vert \bm{h}_i^H\bm{w}_i \vert^2 \\
& & & \beta_{2i} \le  \vert \bm{h}_i^H\bm{w}_C \vert^2 \\
& & & \beta_{3i} \le P_{i} \vert \bm{h}_i^H\bm{w}_i \vert^2 + \vert \bm{h}_i^H\bm{w}_C \vert^2 \\
& & & P_i \ge 0, \;\; 0 \le \alpha_i \le 1, i,j=1,2, i \neq j
\end{aligned}
\end{equation*}

The constraints $\beta_{2i} \le \vert \bm{h}_i^H \bm{w}_C \vert^2$ and $\beta_{3i} \le P_i \vert \bm{h}_i^H\bm{w}_i \vert^2 + \vert \bm{h}_i^h \bm{w}_C \vert^2$ are not convex~\cite{gershman2010convex}. To deal with this, we rewrite the problem into the following form, using the Semidefinite Programming (SDP) framework~\cite{gershman2010convex}. Let $\bm{H}_i = \bm{h}_i \bm{h}_i^H$ and $\bm{W}_C = \bm{w}_C \bm{w}_C^H$. The third term in the objective function can be written as $\Vert \bm{w}_C \Vert^2 = \tr \left( \bm{w}_C \bm{w}_C^H \right) = \tr \left( \bm{W}_C \right)$, where $\tr$ is the trace of a matrix. Also, we can write $\vert \bm{h}_i^H \bm{w}_C \vert^2 = \vert \bm{h}_i^H \bm{w}_C \bm{w}_C^H \bm{h}_i \vert = \tr \left( \bm{h}_i \bm{h}_i^H \bm{w}_C \bm{w}_C^H \right) = \tr \left( \bm{H}_i \bm{W}_C \right)$.
Using the reformulations above, we can convert the problem into the following form:
\begin{equation*}
\begin{aligned}
& \underset{P_i, \bm{W}_C, \alpha_i}{\text{minimize}}
& & P_1 + P_2 + \tr\left( \bm{W}_C \right) \\
& \text{subject to}
& & \beta_{1i} \le P_{i} \vert \bm{h}_i^H\bm{w}_i \vert^2 \\
& & & \beta_{2i} \le \tr\left( \bm{H}_i \bm{W}_C \right) \\
& & & \beta_{3i} \le P_{i} \vert \bm{h}_i^H\bm{w}_i \vert^2 + \tr\left( \bm{H}_i \bm{W}_C \right) \\
& & & \bm{W}_C \succeq 0, \; \rank\left( \bm{W}_C \right)=1 \\
& & & P_i \ge 0, \;\;0 \le \alpha_i \le 1, i=1,2.
\end{aligned}
\end{equation*}

Here, the constraint $\bm{W}_C \succeq 0$ means that matrix $\bm{W}_C$ is positive semidefinite. This problem is not convex because of the rank one constraint. By dropping this constraint, we obtain a lower bound on the value of objective function at the optimal point, since the feasible set is enlargened. This problem can then be solved using tools of SDP. The obtained solution is a lower bound on the transmission power at the BS.
%
%
%
%
\subsection{Finding the Minimum Transmission Power at SBS}\label{sec:transPowerSBS}
As remarked at the end of Sec.~\ref{sec:SystemModel}, the SBS may need to increase its transmission power compared to the wired backhaul case, since the wireless link between SBS$i$ and BS may not be able to support the uplink transmission. Let $\eta_i P_{Si}$ be the required transmission power at SBS$i$ for the wireless backhaul, where $\eta_i \ge 1$.
In the wired backhaul case, the total transmission power at the SBSs is $P_{S1} + P_{S2}$, and in the wireless case, the total transmission power is $\eta_1 P_{S1} + \eta_2 P_{S2}$. Therefore, the extra transmission power in the wireless case compared to the wired case is
\begin{equation}
	(\eta_1 - 1) P_{S1} + (\eta_2 - 1) P_{S2},
\end{equation}
which is to be minimized. 
For simplicity reason, we only treat the special case when $P_{S1} = P_{S2} = P_{S}$, for which the objective function becomes $P_{S} ( \eta_1 + \eta_2 - 2 )$. The minimization of this function is equivalent to minimizing $\eta_1 + \eta_2$.

From our system model, when SBS1 and SBS2 transmit to the BS, the rate region at the BS is a two-sender Single-Input Multiple-Output (SIMO) MAC. The optimization problem is

\begin{equation*}
\begin{aligned}
& \underset{\eta_1, \eta_2}{\text{minimize}}
& & \eta_1 + \eta_2 \\
& \text{subject to}
& & R_{D1} \le \log_2 \left| I_{2M} + \frac{\eta_1 P_{S1} \bm{H}_{1}}{\sigma^2} \right| \\
& & & R_{D2} \le \log_2 \left| I_{2M} + \frac{\eta_2 P_{S2} \bm{H}_{2}}{\sigma^2} \right| \\
& & & R_{D1} + R_{D2} \\ 
& & & \hspace{1.0cm} \le \log_2 \left| I_{2M} + \frac{\eta_1 P_{S1} \bm{H}_{1} + \eta_2 P_{S2} \bm{H}_{2}}{\sigma^2} \right| \\
& & & \eta_1 \ge 1, \; \eta_2 \ge 1.
\end{aligned}
\end{equation*}
where $\log_2 \left| \bm{X} \right| = \log_2 \det \left( \bm{X} \right)$. The extra transmission power is then $P_{S} ( \eta_1 + \eta_2 - 2 )$.

\section{Numerical Results}\label{sec:NumericalResults}
The performance of WEW is demonstrated in this section. 
We assume the channels $\bm{h}_i$ are Rayleigh faded, so the channel coefficients are distributed as $\mathcal{CN}(0,1)$, the zero mean unit variance complex Gaussian distribution. The bandwidth is normalised to 1 Hz. We set $M=1$, so the BS has two antennas. The optimization problems in Sec.~\ref{sec:OptimizationProblem} are solved using the convex optimization software CVX.

For the first example, we compare the proposed scheme WEW with three other schemes. The first one is using only ZF at the BS, while the second one is using only the common beam. We also show the results of partitioning randomly, i.e. for each channel realisation, $\alpha_1$ and $\alpha_2$ are chosen randomly and independently from $\left[ 0, 1 \right]$. 

To allow a meaningful visualization of the results, the uplink and downlink rates are set as simulation parameters. Once their value is fixed, the corresponding SNRs $\gamma_{M_i}$ are derived using  (\ref{eqn:MSdownlinkRateWired}). Furthermore, the MSs are assumed to have the same uplink rate and the same downlink rate.
The results for a fixed uplink rate of 1 bit per second (bps) of both MSs, and varying the downlink rate for both MSs. The results are averaged over 1000 channel realisations. 

\begin{figure}[tbp]
	\centering
		\includegraphics[width=0.45\textwidth]{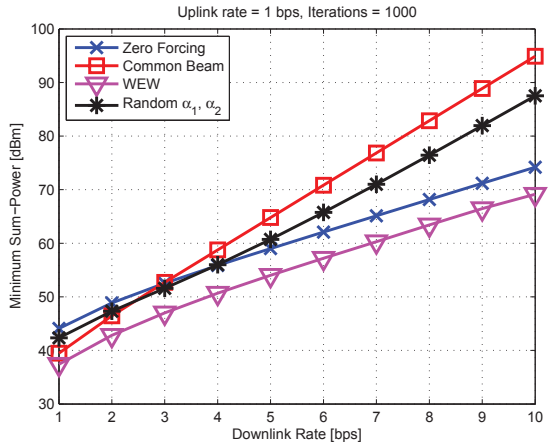}
	\caption{Comparison of minimum transmission power at BS, for the transmission methods considered in this work.}
	\label{fig:minPowerRu1Rd1to10}
\end{figure}

In the first case, shown in Fig.~\ref{fig:minPowerRu1Rd1to10}, we vary the downlink rate between 1 and 10 bps. It is observed that WEW has a performance advantage of about 6 dBm in a large part of the range, compared to ZF and common beam. We also see that the optimization of the splitting factors $\alpha_1$ and $\alpha_2$ results in a gain, compared to just choosing them at random.

Furthermore, we see that there is a crossing point in the downlink rate at about 4 bps. For lower rates, the common beamformer has better performance, while ZF is better for higher rates. Because low rate requirements translate into low SNR requirements, similarly for high rates, the observation is related to the fact that ZF beamforming has an advantage at high SNRs.

\begin{figure}
	\centering
		\includegraphics[width=0.45\textwidth]{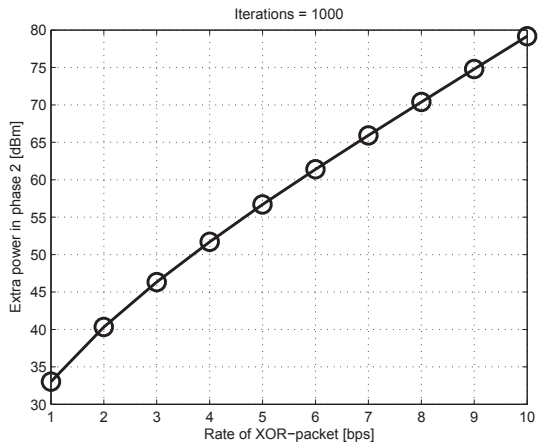}
	\caption{Total additional transmission power required at the SBSs in phase 2.}
	\label{fig:BCpowerRu1Rd1to10}
\end{figure}

In Fig.~\ref{fig:BCpowerRu1Rd1to10}, the results of the optimization problem in Subsec.~\ref{sec:transPowerSBS} is shown, where the rate varies between 1 and 10 bps. Here, it is assumed that the SBS-BS channels are statistically equal to the MS-SBS channels. In a deployment scenario, the SBS would be placed so that these channels were better, which would then lower the required extra power.

\section{Conclusion}\label{sec:Conclusion}
In this paper, we have proposed a transmission scheme to enable wireless backhaul in a cellular communications scenario, consisting of two small-cell MSs, two SBSs and a macro BS. We have proposed a scheme wherein uplink and downlink information flows are treated jointly, using WNC. At the BS, our scheme leverages on transmitting data to the MSs by partitioning it into a public and private part. We formulated an optimization problem to find the minimal transmission power at the BS, subjected to the given rate constraints, which were given from the wired backhaul scenario. Since the initial problem was nonconvex, we used SDP and relaxed the problem, to find a lower bound on the minimal transmission power. Further, we dealt with the additional power required at each SBS, so that the messages could be reliably transmitted over the wireless backhaul link. Numerical results were given to show the performance of the WEW scheme, and we saw that by optimizing the splitting of messages into private and common parts resulted in a performance gain.

\section*{Acknowledgement}
Part of this work has been performed in the framework of the FP7 project ICT-317669 METIS, which is partly funded by the European Union. The authors would like to acknowledge the contributions of their colleagues in METIS, although the views expressed are those of the authors and do not necessarily represent the project.

\appendices{}

\bibliographystyle{ieeetr}
\bibliography{biblio}

\end{document}